# Could scientists use Altmetric.com scores to predict longer term citation counts?[1]

Mike Thelwall, Tamara Nevill, University of Wolverhampton, UK.

Altmetrics from Altmetric.com are widely used by publishers and researchers to give earlier evidence of attention than citation counts. This article assesses whether Altmetric.com scores are reliable early indicators of likely future impact and whether they may also reflect non-scholarly impacts. A preliminary factor analysis suggests that the main altmetric indicator of scholarly impact is Mendeley reader counts, with weaker news, informational and social network discussion/promotion dimensions in some fields. Based on a regression analysis of Altmetric.com data from November 2015 and Scopus citation counts from October 2017 for articles in 30 narrow fields, only Mendeley reader counts are consistent predictors of future citation impact. Most other Altmetric.com scores can help predict future impact in some fields. Overall, the results confirm that early Altmetric.com scores can predict later citation counts, although less well than journal impact factors, and the optimal strategy is to consider both Altmetric.com scores and journal impact factors. Altmetric.com scores can also reflect dimensions of non-scholarly impact in some fields.

## 1    Introduction

Although widely used in support of formal and informal research evaluations, citation counts inevitably lag research by several years due to publication delays. Alternative indicators derived from the web can offset this delay by bypassing the publishing process and recording interest in articles rather than formal citations to them (Neylon & Wu, 2009; Priem, Taraborelli, Groth, & Neylon, 2010). For example, an early career researcher might add altmetrics to her CV before her papers were cited, perhaps also using them to highlight non-scholarly impacts (Piwowar & Priem, 2013). Nevertheless, altmetric scores lack the credibility and theoretical background of citation counts (Fenner, 2014) and evidence is needed of their value to support researchers wishing to use them. Without this, altmetrics might be ignored as irrelevant, given that individual articles may have high scores for spurious reasons. Conversely, they may be given more weight than their values warrant. Altmetric.com is one of the main providers of alternative indicators and its scores are currently (December 2017) visible in the websites of many publishers as well as on its website and displayed as badges by individual researchers (Primary Research Group Inc., 2017). Its scores are therefore widely viewed by academics and may be used for informal judgements of interest in papers as well as in support of research evaluations (e.g., listed on CVs). Evidence of Altmetric.com scores associating with, or predicting, future citation impact would therefore be valuable to many scientists. This article is not attempting to assess whether Altmetric Scores could be used by scientometricians for formal evaluations because indicators that can be manipulated are not suitable for this (Wouters & Costas, 2012). Instead, it investigates whether their informal use by scientists and publishers is likely to give misleading results.

There is evidence that higher altmetric scores associate with higher citation counts for many different indicators, almost always using contemporary rather than future citation counts. Mendeley reader counts have the strongest association with citation counts (Thelwall, 2017), depending on field, and Tweet counts have weak correlations with citation counts (Haustein, Larivière, Thelwall, Amyot, & Peters, 2014). Other alternative indicators tend to be much rarer than these two – often occurring for under 1% of journal articles (Costas, Zahedi, & Wouters, 2015; Thelwall, Haustein, Larivière, & Sugimoto, 2013; Zahedi, Costas, & Wouters, 2014) – and have very weak associations with citation counts (Thelwall, Haustein, Larivière, & Sugimoto, 2013). In terms of predictions, Mendeley reader counts correlate with later citation counts in all fields (Thelwall, submitted), Tweets correlate with later citation counts for articles in one online medical journal (Eysenbach, 2011), and early arXiv preprint downloads correlate with later citation counts for physics (Brody, Harnad, & Carr, 2006). Similarly, statistical regression can predict future usage counts from early usage counts for Nature articles (Wang, Mao, Xu, & Zhang, 2014). Despite these findings, is not known whether early scores can predict later citations for other altmetrics, for Altmetric.com data, or within narrow fields (see RQ2 below).

Some altmetrics may reflect societal impacts, such as health, governmental or commercial benefits. Most Mendeley users (85%) register articles to cite them but some use them for other purposes (professional: 50%; teaching: 25%), suggesting that Mendeley reader counts mainly reflect scholarly impact but also some educational and professional impact (Mohammadi, Thelwall, & Kousha, 2016). Research bloggers discuss articles that may attract general interest (Shema, Bar-Ilan, & Thelwall, 2015), a news-like motivation. Tweeted links to articles seem to be used mainly to share them rather than pointing to alternative types of impact (Robinson-Garcia, Costas, Isett, Melkers, & Hicks, 2017; Thelwall, Tsou, Weingart, Holmberg, & Haustein, 2013). There are disciplinary differences in the way that scholars tweet (Holmberg & Thelwall, 2014), but some believe their tweeting reflects scholarly impact (Priem, & Costello, 2010).

The statistical dimensionality of sets of altmetrics provides indirect evidence that they can reflect multiple impact types. An exploratory factor analysis combined normalised and non-normalised citation indicators and altmetric scores from April 2013 (Mendeley, Twitter, Wikipedia, Delicious bookmarks) for 20,000 randomly selected Web of Science (WoS) publications with DOIs from all fields 2005-2011, finding Delicious and Twitter to associate within a factor and Mendeley, Wikipedia and the citation indicators to associate within another factor (Zahedi, Costas, & Wouters, 2014). A factor analysis of Facebook, Blog, Twitter, Google+, News and Total scores from Altmetric.com in October 2013 for WoS publications with DOIs and published July-December 2011, together with five normalised and non-normalised citation indicators (Costas, Zahedi, & Wouters, 2015) found News and Blogs to be in the same factor, with Facebook, Twitter, and Google+ being in a second factor (with Total, which incorporates them), separate from the bibliometric measures. This suggests a news-related dimension and a social web discussion dimension. An analysis of 4 citation and 18 altmetric scores (viewing, saving, social web discussion, media – some overlapping, such as monthly and total view counts) gathered from PLOS in October 2015 for 32,862 PLoS ONE papers published in 2014 found a citation factor and a separate view count factor that included Mendeley reader counts (Buttliere & Buder, 2017). This paper log transformed the data to reduce skewing. A principal components analysis of multiple publication output, webometric and achievement indicators for 57 social sciences and humanities scholars in Taiwan found web presence indicators to be in a separate factor to publication outputs (Chen, Tang, Wang, & Hsiang, 2015). These analyses have used multidisciplinary data sets and so the factors found could be due to disciplinary differences and they do not reveal if there are disciplinary differences in the importance of the factors, if they reflect societal impact (see RQ1 below).

This paper investigates whether the scores provided by Altmetric.com would be misleading indicators of future citation impact and whether there is statistical evidence from exploratory factor analysis that they reflect non-scholarly types of impact. This is the first diachronic analysis of Altmetric.com scores. Previous investigations of Altmetric.com scores have not attempted to compare early Almetric.com Scores with later citation counts. This is an important omission because this is an obvious use for them by working scientists. Exploratory factor analysis (EFA) uncovers evidence of hidden variables (e.g., factors like scholarly impact, societal impact, newsworthiness, interest) that influence measured variables (e.g., altmetric scores) to explain the correlations between the measured variables (Cudeck, 2000). If EFA reveals a factor influencing multiple variables then this is evidence of a systematic hidden relationship but not proof of its existence or meaning and so it is a weak method. The following research questions drive the study.

1. Are there systematic relationships between Altmetric.com scores for academic papers within narrow fields that would be consistent with them reflecting different dimensions of impact?
2. Can Altmetric.com scores from close to the publication date of an article predict longer term citation counts?
3. Are there disciplinary differences in the answers to the above questions?

# 2   Methods

The first part of the research design was to investigate relationships between variables using correlations and factor analysis on relatively mature Altmetric.com scores for a range of narrow fields to give increased statistical power to identify hidden factors in comparison to a less mature dataset. The second part of the research design was to predict current citation counts from earlier Altmetic.com scores for the same fields.

## 2.1   Data

Altmetric scores were obtained from Altmetric.com (Adie & Roe, 2013) for their complete dataset under their research data sharing programme. The most recent dataset, October 2017, was used for the study of patterns between scores, since this would have the most comprehensive data. An Altmetric.com dataset from November 2015 was selected to give older data (2 years) to investigate whether Altmetric.com scores in the first year of publication could be used to predict later citation counts.

Scopus categorises publications into 310 narrow fields, grouped into 27 broad fields. Each narrow field is given a four-digit numerical code, with the first two digits being the broad field code. Narrow field codes are given out consecutively for each broad field. For example, Electrochemistry (1603) and Inorganic Chemistry (1604) are consecutive members of the Chemistry broad field (16). Some broad fields contain few narrow fields (e.g., Economics, Econometrics and Finance: 3) and some have many (e.g., Medicine: 48). Thirty narrow Scopus fields were chosen for wide coverage of different areas to investigate disciplinary differences. To be systematic, the second narrow category of each broad Scopus category was chosen. The 22nd and 42nd fields were also selected when there were enough narrow fields. This gives greater representation to large Scopus broad fields. *Dental*

*Assisting* had no results in one of the years investigated and was replaced with *Oral Surgery* from the same broad field, *Dentistry*.

Scopus citation counts were obtained for all documents of type journal article from the 30 narrow fields for 2013 and 2015. The year 2013 was selected to give a mature data set (4 years old) for the correlations and factor analysis. The year 2015 was selected to be contemporary with the November 2015 dataset to investigate citation predictions, as described below. The citation counts for the chronologically first and last 5000 documents of type journal article in each year were extracted from Scopus in October 2017. For some fields, the results were not comprehensive and would omit papers published in the middle of the year, a Scopus data collection limitation. This should not greatly influence the results because each year is time-balanced.

The Altmetric.com and Scopus datasets were merged based on article DOIs, with articles lacking a DOI being discarded. This step was taken in preference to article metadata matching heuristics because this could introduce errors.

The citation counts and Altmetric.com scores were all transformed with the formula $ln(1 + x)$ before processing to reduce skewing (following the example of: Buttliere & Buder, 2017). This transformation is necessary for linear regression involving citations and Mendeley readers (Thelwall & Wilson, 2016) and probably also for almost all altmetrics giving the high general level of skewing (Buttliere & Buder, 2017).

Articles in Scopus without a matching Altmetric.com record were discarded. Keeping the data would introduce errors since Altmetric.com collects data from Mendeley only for articles that it has found through other methods and so its Mendeley data is incomplete. This step also removes the problem for the November 2015 Altmetric data set applied to the 2015 Scopus articles of including unpublished articles since these would presumably have no Altmetric scores and would therefore be discarded.

## 2.2 Association analysis

Spearman correlations were calculated as a basic measure of association between pairs of variables. They are useful to assess whether there could be an underlying relationship that would allow a claim that the indicators reflected a common type of impact (Sud & Thelwall, 2014). Citation counts and alternative indicators are a discrete data type, often with many zeros. This reduces the underlying association strength and so correlation coefficients are not directly comparable between scores (Thelwall, 2016a). Nevertheless, the direction of commonality can be identified – if higher ranked articles with one score tend to be highly ranked with another. For this analysis, the three weakest indicators with almost all zeros (Connotea, F1000, Videos) were removed since their scores were non-zero for too few articles in each field and year to be informative. They were also zero for all articles in some fields, which prevents factor analysis from working.

Spearman correlations can obscure multidimensional relationships between variables, especially if there is one strong relationship. Exploratory factor analysis was therefore used to identify multidimensionality that could be due to additional impact types. The standard expectation maximisation method was not used to fit the factors because the variables had skewness and kurtosis above the recommended ranges (West, Finch, & Curran, 1995) despite the skewness reduction through the log transformation. Instead, principal axis factoring was used with the *fa* command in the *psych* package of the statistical software R. This method is safest for data that is not normally distributed (Fabrigar, Wegener, MacCallum, & Strahan, 1999). Promax rotation was used to make the factors easier to interpret (following: Buttliere & Buder, 2017). It is an oblique rotation method and is preferable to orthogonal methods like varimax because there is no reason to believe that the main influences (factors or latent variables) would be uncorrelated.

For each of the 29 factor analyses (one factor analysis failed due to the absence of non-zero scores on one variable), the number of factors to be extracted was determined by the parallel analysis technique (R command *fa.parallel*) which reports the number of factors that would account for more variance that expected by chance. In many cases the cut-off point was close to the margin and in some cases the parallel analysis produced errors. Thus, in practice, the number of factors decision was sometimes not clear cut. Factor loadings above 0.3 were included so that the factors could be interpreted inclusively. The raw data, factor analysis results, factor loading diagrams and parallel analysis diagrams are saved in a file associated with this paper and only summary results are reported here.

## 2.3 Prediction analysis

Ordinary least squares linear regression was used to generate a prediction formula for future Scopus citation counts (October 2017) from November 2015 altmetric data for articles published in 2015, excluding articles with no positive Altmetric.com score. The regression is therefore an attempt to predict citation counts two years in the future from Altmetric.com scores from the publication year. This is not strictly speaking prediction since the future citation counts are used to generate the equation. Instead it is an attempt to build a prediction formula together with statistical evidence that the formula would be effective in predicting future citations if the system remained unchanged.

Linear regression on the log transformed citation counts is more effective than discrete regression methods for citation data (Thelwall & Wilson, 2014; Thelwall, 2016b). The Enter method was used for the linear regression, which fits all variables in one attempt, rather than adding them sequentially. This ensures consistency between the 30 different regression equations. Statistically significant coefficients were reported as the potentially useful future citation impact scores. The regressions were repeated without Mendeley reader counts because these are not transparent (cannot be verified) and so Altmetric.com categorises them as a lower value score.

# 3  Results

The R code used and complete sets of results from the factor analysis (factor loadings, parallel analysis diagrams) are available in the online appendix.

## 3.1  Relationships between variables

Most indicators have weak positive correlations with each other, on average (median: Table 1). The main exception is the strong average correlation between Mendeley readers and Scopus citation (0.624). The second strongest median correlation is between the similar News and Blogs (0.246). All indicators have positive correlations with Scopus citation counts (and Mendeley reader counts), suggesting that they all tend to relate to scholarly communication to some extent, even if only at a very minor level. On average, there was essentially no relationship between the number of tweets and Facebook posts about articles and also no relationship between them and future citation counts.

Table 1. Median Spearman correlations between pairs of variables for each narrow field (n=30) for articles published in 2013 with citation counts and Altmetric.com scores from October 2017.

| Median* | Scopus | Mendeley | Blogs | News | Google+ | Twitter | Facebk | Wiki. |
|---|---|---|---|---|---|---|---|---|
| **Scopus** | 1.000 | **0.624** | 0.105 | 0.113 | 0.048 | 0.084 | 0.053 | 0.051 |
| **Mendeley** | **0.624** | 1.000 | 0.137 | 0.122 | 0.090 | 0.148 | 0.072 | 0.051 |
| **Blogs** | 0.105 | 0.137 | 1.000 | **0.246** | 0.104 | 0.026 | 0.049 | 0.023 |
| **News** | 0.113 | 0.122 | **0.246** | 1.000 | 0.101 | 0.028 | 0.038 | 0.023 |
| **Google+** | 0.048 | 0.090 | **0.104** | 0.101 | 1.000 | 0.099 | 0.098 | 0.018 |
| **Twitter** | 0.084 | **0.148** | 0.026 | 0.028 | 0.099 | 1.000 | -0.007 | -0.082 |
| **Facebook** | 0.053 | 0.072 | 0.049 | 0.038 | **0.098** | -0.007 | 1.000 | -0.016 |
| **Wikipedia** | **0.051** | **0.051** | 0.023 | 0.023 | 0.018 | -0.082 | -0.016 | 1.000 |

* Bold correlations are the strongest for the row.

A factor analysis was possible for 29 of the fields (Table 2). It was not possible for Research and Theory because there were no Wikipedia citations. Scopus citations and Mendeley readers dominated the results, always appearing jointly in a factor that almost always did not include another score. This factor presumably accounts for traditional scholarly impact so the remaining factors suggest other types of attention or impact.

Blogs and news formed a separate factor 12 out of 29 times (41%) and the two were part of a larger factor a further 8 times (28%). Thus, there is evidence of a newsworthiness factor in two fifths of all fields, and perhaps more if other altmetrics contribute to it.

Wikipedia occurred with other variables in a factor only twice (Histology; Urban Studies) suggesting that it partly reflects a unique type of impact, or nothing. If it exists, this might be thought of as informational impact.

The associations between Google+, Twitter and Facebook with each other and with News/Blogs by field, without a clear trend. All five were in the same factor in five fields, and a possible interpretation is that newsworthiness drives social media coverage in these fields. The first three sites were in a single factor away from the other variables in three fields, where they might represent social web discussions. Negative associations between Twitter and Facebook in three factors might represent areas where different subfields or journals use one or the other.

To check whether the results could be due to the rotation method, the factor analyses were repeated with oblimin (oblique) rotation. This made minor changes to the relationships between the factors and altmetrics for six of the results (History; Information Systems & Management; Histology; Equine; Oral Surgery; Chiropractics) but these would not alter the overall conclusions. For example, for Chiropractics the factors were identical between oblimin and promax except that promax had one additional variable loading on one of the factors: Mendeley with the minimum value taken as significant (0.3).

Table 2. Factors from factor analyses in each field in order of importance (Scopus publications from 2013, Altmetric.com scores and Scopus citation counts from October 2017). Three altmetrics with mainly zero scores were excluded. Shading and repeated numbers are for visual effect. Negatives indicate negative factor loadings. For example, two factors were extracted from History. The most important, factor 1, included Scopus citations and Mendeley readers. The second most important, factor 2, consisted of blogs and news scores.

| Field | Cited | Mend | Blogs | News | Goog+ | Twitter | Facebk | Wiki |
|---|---|---|---|---|---|---|---|---|
| Agronomy and Crop Science | 11111 | 11111 | 4 | 4 | | | 33 | 2222 |
| History | 11111 | 11111 | 2222 | 2222 | | | | |
| Aging | 11111 | 11111 | 2222 | 2222 | 33 | 33 | 33 | |
| Accounting | 11111 | 11111 | 33 | 33 | | | 2222 | |
| Bioengineering | 11111 | 11111 | 4 | 4 | 2222 | 2222 | 5 | |
| Analytical Chemistry | 11111 | 11111 | 33 | 33 | | 2222 | | |
| Artificial Intelligence | 11111 | 11111 | 2222 | 2222 | | 4 | | 33 |
| Information Systems & Manag. | 11111 | 11111 | 2222 | 2222 | 222233 | 33 | 2222 | 4 |
| Atmospheric Science | 11111 | 11111 | 2222 | 2222 | 2222 | 2222 | 2222 | 33 |
| Economics and Econometrics | 11111 | 11111 | 4 | 4 | 33 | 33 | 33 | 2222 |
| Energy Eng. & Power Tech. | 2222 | 2222 | 33 | | | 11111 | -11111 | |
| Aerospace Engineering | 11111 | 11111 | 6 | 4 | 5 | 6 | 33 | 2222 |
| Ecological Modeling | 2222 | 2222 | 11111 | 11111 | 11111 | 11111 | 11111 | 33 |
| App. Microbiology & Biotech. | 11111 | 11111 | 2222 | 2222 | 4 | 2222 | 4 | 33 |
| Biomaterials | 11111 | 11111 | 2222 | 2222 | | 33 | | |
| Algebra and Number Theory | 2222 | 2222 | 33 | | | -11111 | 11111 | |
| Anatomy | 2222 | 2222 | 11111 | 11111 | | | 4 | 33 |
| Histology | 2222 | 2222 | 11111 | 11111 | 11111 | 33 | 33 | 11111 |
| Rehabilitation | 2222 | 2222 | 4 | 4 | 11111 | | 11111 | 33 |
| Behavioral Neuroscience | 2222 | 2222 | 11111 | 11111 | 11111 | 11111 | 11111 | 33 |
| Adv. & Specialized Nursing Research and Theory | 2222 | 2222 | 11111 | 11111 | 11111 | 11111 | 11111 | 33 |
| Drug Discovery | 11111 | 11111 | | 5 | | 4 | 2222 | 33 |
| Acoustics and Ultrasonics | 11111 | 11111 | 2222 | 2222 | 33 | 4 | 2222 | |
| Applied Psychology | 11111 | 11111 | 33 | 33 | 4 | 4 | 4 | 2222 |
| Archeology | 11111 | 11111 | 2222 | 2222 | 2222 | 2222 | 2222 | 33 |
| Urban Studies | 11111 | 11111 | 33 | 33 | | | 2222 | 33 |
| Equine | 11111 | 11122 | | | 33 | 2222 | 22233 | |
| Oral Surgery | 11111 | 11111 | | | | -2222,4 | 2222 | |
| Chiropractics | 11111 | 11122 | 33 | | 33 | 2222 | 2222 | |

## 3.2 Prediction

The primary prediction task is to assess whether longer term Scopus citation counts (October 2017) can be predicted from early Altmetric.com scores (November 2015) for articles in the same year (2015). From a basic examination of the data (Table 3), the low scores and percentage cited are clear for Connotea, F1000 and Video, indicating that these are rare enough to have little value for individual articles.

Removing articles with zero Altmetric.com scores increases the median correlations with Scopus citations substantially for Mendeley (0.247 to 0.391), CiteULike (0.069 to 0.081) and F1000 (0.011 to 0.027) and substantially decreases it for Twitter (0.198 to 0.135) and Facebook (0.105 to 0.072). The increase is expected for Mendeley because many of the articles in the raw data set with no Mendeley readers might not have been queried via the Mendeley API. The increase for some non-Mendeley altmetrics suggests that not all their zero scores are genuine zeros. Nevertheless, it is better to exclude articles without any positive altmetric scores from the results because invalid data is worse than truncated data. In practice, authors should be cautious about interpreting zero scores as evidence that their article has received no online attention (occurring for 73% of articles in November 2015 for the 2015 data set) since it may have been mentioned in ways that Altmetric.com does not track.

Note that Twitter is numerically dominant compared to the other indicators, so "Articles with a positive Altmetric.com score" is almost synonymous with "Articles that have been tweeted" because 93.7% of articles with an Altmetric.com score have been tweeted (Table 3).

Table 3. Median descriptive statistics for the 30 fields for articles published in 2015 calculated separately. The results are presented with (All) or without (No 0) articles with zero scores for all altmetrics. The Altmetric.com scores are from November 2015 and the future citation counts are from October 2017.

|  | Articles with a positive Altmetric.com score | | | All articles | | |
|---|---|---|---|---|---|---|
| Indicator | Geometric mean | Correlation with 2017 citations | Percent nonzero | Geometric mean | Correlation with 2017 citations | Percent nonzero |
| Scopus 2017 | 3.28 | 1.000 | 86.2% | 2.24 | 1.000 | 76.4% |
| Mendeley | 1.91 | 0.391 | 70.4% | 0.40 | 0.247 | 23.5% |
| CiteULike | 0.02 | 0.081 | 2.6% | 0.01 | 0.069 | 0.7% |
| Connotea | 0.00 | -0.006 | 0.0% | 0.00 | -0.002 | 0.0% |
| Blogs | 0.04 | 0.089 | 5.1% | 0.01 | 0.083 | 1.3% |
| News | 0.05 | 0.074 | 4.3% | 0.01 | 0.079 | 1.3% |
| Google+ | 0.01 | 0.054 | 1.6% | 0.01 | 0.047 | 0.5% |
| Twitter | 2.10 | 0.135 | 93.7% | 0.40 | 0.198 | 26.8% |
| Facebook | 0.20 | 0.072 | 21.6% | 0.06 | 0.105 | 6.4% |
| Video | 0.00 | 0.008 | 0.0% | 0.00 | 0.007 | 0.0% |
| Wikipedia | 0.01 | 0.039 | 0.7% | 0.00 | 0.034 | 0.2% |
| F1000 | 0.00 | 0.027 | 0.0% | 0.00 | 0.011 | 0.0% |

Mendeley is the strongest predictor of future citations (Table 4). It is positive and statistically significant in all regression equations. Although CiteULike is rare (Table 3), when statistically significant it has a substantial positive coefficient (Table 4). These results seem reasonable since both Mendeley and CiteULike are tools to support academic citing.

Video citation score coefficients are negative in the three cases where they are significant, but these are for small numbers. In Economics and Econometrics, only one paper, "The Academic Entrepreneur: A Biographical Sketch of Ian C. MacMillan's Contributions to Establishing the Field of Entrepreneurship" has a video citation (from a YouTube video[2] made by the [brave] author to support her article) and this is uncited despite having 10 Mendeley readers, 6 tweets, a CiteULike user and a blog post. Thus, the rareness of Video citations allows them here to "compensate" for the other high scores when predicting a zero citation count (i.e., a marginal effect in statistical terminology). Although a single article is too few to decide from, it is possible that video citations associate with non-scholarly impact, such as educational value.

The remaining Altmetrics.com scores are more ambiguous, being non-significant in most equations (except for Twitter), mostly having low values (except for Wikipedia) and having at least one negative coefficient. Since the coefficients are judged for statistical significance at p=0.05 and there are multiple coefficients to test, a few anomalous results could be expected. If the negative cases are not all due to statistical anomalies related to the statistical significance level used, then alternative explanations are possible. Non-independence could produce spurious results, such as an active editor of a low impact journal tweeting all its articles or a blogger focusing on a low-citation specialism. Alternatively, negative coefficients can be a marginal effect: a high correlation with another independent variable that is a better prediction of citation counts, combined with a relatively stronger association with non-citation impact.

---

[2] https://www.youtube.com/watch?v=Iepvf-kgHuw

Table 4. Statistically significant linear regression coefficients. The Altmetric.com scores are from November 2015 and the future citation counts are from October 2017 for Scopus articles from 2015, **excluding** articles with no positive Altmetric.com scores. All Connotea and F1000 coefficients were non-significant. Negative coefficients are highlighted.

| Field | Mend. | CiteU | Blogs | News | G+ | Twit | Face | Vid | Wiki | n | R²* |
|---|---|---|---|---|---|---|---|---|---|---|---|
| Agronomy & Crop Sci | 0.38 | 0.51 | -0.36 | 0.24 | | 0.23 | | | | 1281 | 24% |
| History | 0.43 | | 0.23 | | | | 0.09 | | 0.39 | 1770 | 25% |
| Aging | 0.38 | | | 0.13 | | 0.14 | | | | 1561 | 25% |
| Accounting | 0.36 | 1.17 | | | -0.72 | | | | | 566 | 24% |
| Bioengineering | 0.27 | 0.30 | | 0.25 | | 0.24 | | | | 2278 | 22% |
| Analytical Chemistry | 0.31 | | | | | 0.25 | | | | 1983 | 13% |
| Artificial Intelligence | 0.43 | | | | | -0.15 | -0.19 | | 1.22 | 962 | 21% |
| Info. Systems & Man | 0.39 | | | | 0.44 | | 0.36 | | 1.72 | 517 | 27% |
| Atmospheric Science | 0.30 | | | 0.30 | | 0.18 | -0.09 | | | 3473 | 16% |
| Econ & Econo | 0.37 | 0.64 | | 0.15 | | 0.10 | | -3.17 | | 2098 | 23% |
| Energy Eng & Power | 0.46 | | | | | | | | | 226 | 21% |
| Aerospace Eng. | 0.32 | | | | | | | | | 320 | 11% |
| Ecological Modeling | 0.31 | | | 0.26 | | 0.09 | -0.43 | | | 784 | 18% |
| App Microb & Biotec. | 0.36 | 0.41 | | 0.26 | | 0.23 | | | 0.50 | 3259 | 34% |
| Biomaterials | 0.38 | | | | | | | | | 2370 | 18% |
| Algebra & Numb Th | 0.39 | | | | | | | | 0.98 | 121 | 14% |
| Anatomy | 0.38 | | | -0.20 | | 0.22 | | | | 1410 | 26% |
| Histology | 0.42 | 0.46 | | | | 0.14 | | | | 1832 | 24% |
| Rehabilitation | 0.36 | | | 0.20 | 0.18 | 0.17 | -0.07 | | 0.82 | 2767 | 22% |
| Behavioral Neurosci | 0.32 | 0.19 | | 0.12 | | | | | 0.63 | 3643 | 18% |
| Adv & Spec. Nursing | 0.55 | 0.65 | 0.49 | 0.26 | -0.59 | 0.36 | | | | 1773 | 42% |
| Research & Theory | 0.36 | | | | | 0.21 | | | | 144 | 18% |
| Drug Discovery | 0.24 | | 0.65 | 0.25 | | 0.18 | | | | 1896 | 10% |
| Acoustics & Ultra | 0.37 | | | | | -0.15 | 0.34 | | | 430 | 17% |
| Applied Psych. | 0.34 | | | 0.13 | | 0.15 | 0.09 | -1.66 | | 3509 | 23% |
| Archeology | 0.36 | 0.91 | | | | | | | | 903 | 21% |
| Urban Studies | 0.41 | | | 0.29 | | 0.11 | | -2.33 | -1.73 | 950 | 28% |
| Equine | 0.15 | | | | | | | | | 267 | 6% |
| Oral Surgery | 0.38 | | 0.84 | | | 0.17 | | | | 1117 | 18% |
| Chiropractics | 0.26 | | 0.61 | | | | | | | 153 | 27% |

*All regression fits statistically significant at p=0.001 except Equine (p=0.060).

The results do not change much if articles without any positive Altmetric.com scores are included (Table 5).

Table 5. Statistically significant linear regression coefficients. The Altmetric.com scores are from November 2015 and the future citation counts are from October 2017 for Scopus articles from 2015, **including** articles with no positive Altmetric.com scores. All F1000 coefficients were non-significant. Negative coefficients are highlighted.

| Field | Mend | CiteU | Con | Blogs | News | G+ | Twit | Face | Vid | Wiki | n | $R^{2*}$ |
|---|---|---|---|---|---|---|---|---|---|---|---|---|
| Agronomy & Crop Sci | 0.37 | 0.49 | | -0.39 | | | 0.19 | -0.11 | | | 8556 | 7% |
| History | 0.43 | | | 0.23 | | | 0.05 | 0.10 | | 0.39 | 6221 | 15% |
| Aging | 0.35 | | | | | | 0.08 | | | | 2364 | 22% |
| Accounting | 0.34 | 1.18 | | | | -0.70 | | | | | 3012 | 11% |
| Bioengineering | 0.28 | 0.24 | | | 0.24 | | 0.29 | | | | 9277 | 13% |
| Analytical Chemistry | 0.29 | | | | | | 0.18 | | | | 9564 | 5% |
| Artificial Intelligence | 0.41 | | | | | | -0.21 | -0.24 | | 1.11 | 9488 | 3% |
| Info. Systems & Man | 0.29 | | | | | 0.55 | -0.14 | | | 1.50 | 3395 | 4% |
| Atmospheric Science | 0.25 | | | | 0.31 | | 0.07 | | | | 9556 | 7% |
| Econ & Econo | 0.34 | 0.65 | | | 0.15 | | | | -3.12 | | 7764 | 10% |
| Energy Eng & Power | 0.48 | | | | | | | | | | 8752 | 1% |
| Aerospace Eng. | 0.30 | | | | | | | | | | 7887 | 1% |
| Ecological Modeling | 0.27 | | | | | | | -0.41 | | | 2691 | 8% |
| App Microb & Biotec | 0.32 | 0.54 | | | 0.28 | | 0.13 | | | 0.49 | 8587 | 20% |
| Biomaterials | 0.36 | | | | | | | | | | 9764 | 6% |
| Algebra & Numb Th | 0.39 | | | -2.57 | | | | | | 1.01 | 5434 | 1% |
| Anatomy | 0.39 | | | | -0.20 | | 0.24 | | | | 3057 | 32% |
| Histology | 0.41 | 0.48 | | | | | 0.10 | | | | 3903 | 18% |
| Rehabilitation | 0.33 | | -2.91 | | 0.22 | 0.19 | 0.13 | -0.07 | | 0.84 | 6116 | 18% |
| Behavioral Neurosci | 0.27 | 0.20 | -2.62 | | 0.11 | 0.22 | -0.06 | | | 0.64 | 5685 | 13% |
| Adv & Spec. Nursing | 0.52 | 0.74 | | 0.50 | 0.31 | -0.43 | 0.26 | | | | 2328 | 40% |
| Research & Theory | 0.32 | | | | | | 0.16 | | | | 334 | 17% |
| Drug Discovery | 0.27 | | | 0.65 | 0.29 | | 0.30 | 0.19 | | | 7378 | 9% |
| Acoustics & Ultra | 0.36 | | | | | | -0.19 | | | | 5227 | 2% |
| Applied Psych. | 0.30 | | | | 0.14 | | 0.07 | 0.09 | | | 6908 | 19% |
| Archeology | 0.37 | 0.92 | | | | | 0.07 | | | | 2659 | 17% |
| Urban Studies | 0.36 | | | | 0.29 | | | | -2.51 | -2.00 | 2644 | 14% |
| Equine | 0.11 | | | | | | | | | | 759 | 2% |
| Oral Surgery | 0.33 | | | 0.77 | | | | | | | 3562 | 7% |
| Chiropractics | 0.25 | | | 0.62 | | | | | | | 205 | 26% |

*All regression fits statistically significant at p=0.001 except Equine (p=0.065).

If Mendeley reader counts are excluded (as they are from the main Altmetric.com score) then there are half as many negative coefficients (7 instead of 14) and there are more statistically significant coeffects from the remaining altmetrics (83 instead of 71) (Table 6).

Table 6. Statistically significant linear regression coefficients, **excluding Mendeley**. The citation counts are from October 2017 and the Altmetric.com scores for the same articles are from November 2015 for Scopus articles from 2015, **excluding** articles with no positive Altmetric.com scores. All F1000 coefficients were non-significant. Negative coefficients are highlighted.

| Field | CiteU | Con | Blogs | News | G+ | Twit | Face | Vid | Wiki | n | $R^{2*}$ |
|---|---|---|---|---|---|---|---|---|---|---|---|
| Agronomy & Crop Sci | 0.66 | | | | | 0.37 | | | | 1281 | 9% |
| History | | | 0.28 | | | 0.10 | 0.19 | | 0.51 | 1769 | 5% |
| Aging | 0.45 | | | 0.15 | | 0.19 | 0.15 | | 0.35 | 1561 | 8% |
| Accounting | 1.11 | | | | | | 0.50 | | | 566 | 4% |
| Bioengineering | 0.55 | | 0.25 | 0.27 | | 0.23 | | | | 2277 | 14% |
| Analytical Chemistry | 0.63 | | | | | 0.29 | | | | 1980 | 4% |
| Artificial Intelligence | | | | | | | | | 0.95 | 958 | 1% |
| Info. Systems & Man | | | | | 0.55 | | 0.51 | | 1.65 | 516 | 6% |
| Atmospheric Science | | | 0.18 | 0.27 | | 0.23 | | | 0.85 | 3470 | 6% |
| Econ & Econo | 0.76 | | | 0.25 | | 0.15 | -2.76 | | | 2097 | 4% |
| Energy Eng & Power | | | | | | | | | 1.67 | 224 | 4% |
| Aerospace Eng. | | | | | | | | | | 318 | 4% |
| Ecological Modeling | | | | | | 0.20 | -0.34 | | | 782 | 4% |
| App Microb & Biotec | 0.69 | | | 0.27 | | 0.29 | 0.17 | | | 3256 | 21% |
| Biomaterials | | | 0.26 | | | | 0.16 | | | 2368 | 2% |
| Algebra & Numb Th | | | | | | | | | 1.30 | 120 | 6% |
| Anatomy | 0.44 | | | -0.28 | | 0.31 | | | | 1408 | 9% |
| Histology | 0.83 | | | | | 0.21 | | | | 1832 | 6% |
| Rehabilitation | 0.44 | -2.00 | | 0.23 | 0.18 | 0.26 | | | 1.13 | 2767 | 8% |
| Behavioral Neurosci | 0.42 | -2.61 | | | 0.22 | 0.06 | 0.13 | | 0.60 | 3643 | 5% |
| Adv & Spec. Nursing | 1.22 | | 0.69 | 0.28 | -0.72 | 0.45 | 0.10 | | | 1773 | 26% |
| Research & Theory | | | | | | 0.18 | | | | 144 | 5% |
| Drug Discovery | | | 0.65 | 0.35 | | 0.21 | 0.27 | | 0.66 | 1894 | 6% |
| Acoustics & Ultra | | | | | | | 0.54 | | | 430 | 4% |
| Applied Psych. | 0.31 | | | 0.15 | | 0.21 | 0.18 | -1.93 | | 3509 | 8% |
| Archeology | 0.97 | | | | | | | | | 903 | 7% |
| Urban Studies | | | | 0.31 | | 0.24 | | | | 950 | 7% |
| Equine | | | | | | | | | | 267 | 2% |
| Oral Surgery | | | 1.24 | | | 0.23 | 0.15 | | | 1116 | 5% |
| Chiropractics | | | | | | 0.19 | | | | 153 | 17% |

*All regression fits statistically significant at p=0.001 except Accounting (p=0.001), Artificial Intelligence (p=0.224), Energy Engineering (p=0.220), Aerospace (p=0.127), Algebra (p=0.136), Research (p=0.353), Acoustics (p=0.032), Equine (p=0.515).

## 3.3 Prediction accuracy

The main regression equations for predicting future (2017) citation counts from prior (2015) altmetric data for articles from 2015 with at least one non-zero altmetric score (Table 4) are all statistically significant fits except for Equine. Thus, in 29 out of 30 fields, predicting future citation counts from current Altmetric.com scores gives better results than simple guessing.

The predictions are not strong, however. Excluding Equine, the percentage of variance explained by the (log transformed) data varies between 10% (Drug Discovery) and 42% (Advanced and Specialised Nursing), with the average being 22%. Thus Altmetric.com scores can account for about a fifth of the variability in citation counts for articles published up to two years later. For extra background context, if Altmetric.com scores from 2015 for 2013 papers (i.e., two years after publication) are used to predict 2017 citations (four years after publication) using the same linear regressions the average variance explained rises to 39% and all regressions are statistically significant. In the extreme non-prediction case of 2017 citation counts estimated (no longer predicted) from November 2017 Altmetric scores for papers from 2013, 40% of the variance is explained by the regression equation, on average (but only 12% if articles without positive Altmetric.com scores are kept).

The prediction equation for the strongest case, Advanced and Specialised Nursing (Table 4) including only statistically significant coefficients is illustrated here. The purpose of the current paper is not to derive prediction equations but to assess whether the current informal researcher practice consulting Altmetric.com scores for longer term citation prediction is reasonable. The equation is therefore presented to help understand the methods rather than as a recommendation.

$$
\begin{aligned}
ln(1 + Citations) &= 0.38 + 0.55 * ln(1 + Mendeley) + 0.65 * ln(1 + CiteULike) + 0.49 \\
&\quad * ln(1 + Blogs) + 0.26 * ln(1 + News) - 0.59 * ln(1 + GooglePlus) \\
&\quad + 0.36 * ln(1 + Tweets)
\end{aligned}
$$

Or

$$
\begin{aligned}
Citations &= e^{0.38}(1 + Mendeley)^{0.55}(1 + CiteULike)^{0.65}(1 + Blogs)^{0.49}(1 \\
&\quad + News)^{0.26}(1 + GooglePlus)^{-0.59}(1 + Tweets)^{0.36} - 1
\end{aligned}
$$

## 4   Discussion

The results are limited by the choice of subject areas and years. Although the coverage of fields is wide, there may be other subject areas for which Altmetric.com scores are substantially less, or more, powerful for predicting future citations. The factor analysis results are limited by the nature of the data. Even after removing three altmetrics, some of the remaining variables contained many zeros, which makes the factor fitting vulnerable to the properties of a small number of articles. The same is true for the linear regression and is more important because it assumes normally distributed data. Although ordinary least squares regression can work well on non-normal data and the scores were log transformed to reduce skewing, the reported $p$ values for coefficients are unsafe. From a different perspective, the inclusion of articles with ages varying from 1 day to 11 months in the November 2015 Altmetric.com/Scopus 2015 combined dataset undermines the power of the main regression analysis reported and so the underlying predictive power of Altmetric.com scores is probably higher than reported above.

Another limitation of the method used here is that regression is not prediction in the pure sense. To test the predictive power of a regression formula, the same formula would need to be applied to data from a subsequent year to see if it still worked. Although the statistical significance of the regression coefficients suggests that this would be the case, it is possible that system changes between years would render a formula from one year obsolete the following year. Large system changes seem unlikely, however.

Exploratory factor analysis does not provide strong statistical evidence because it does not explicitly test a hypothesis and because alternative mathematical models could explain the data in different ways (e.g., unrotated factor loadings), although agreement across multiple related EFAs increases the strength of the cumulative evidence. Finally, the factor analysis and linear regression assume that the independent variables are independent but it is known that there are common influences, so the independence assumption is not fully true. For example, author nationality affects both citation counts and Mendeley reader scores (Thelwall & Maflahi, 2015).

The factor analyses confirm prior studies that have found Mendeley to be the altmetric with the strongest association with citation counts (e.g., Haustein, Larivière, Thelwall, Amyot, & Peters, 2014; Thelwall, Haustein, Larivière, & Sugimoto, 2013) and to have a strong correlation with citation counts in almost all fields (Thelwall, 2017). They extend prior knowledge (Zahedi, Costas, & Wouters, 2014) by showing that a range of other altmetrics (Twitter, Google+, Facebook, Blogs, News) tend to at least partly reflect one or more different dimensions of interest and that the results hold for narrow fields and when only comparable scores are included in the factor analysis. Blogs and News are often found together as a newsworthiness dimension, confirming a prior study if general interest is interpreted as newsworthiness (Shema, Bar-Ilan, & Thelwall, 2015). Wikipedia seems to be a unique altmetric, despite prior evidence of positive (low) correlations with citation counts in many fields (Kousha & Thelwall, 2017). In contrast, Twitter, Google+ and Facebook seem to vary their meanings by field without a common pattern. This partly contradicts scholars that believe that their tweeting reflects scholarly impact (interviews in: Priem & Costello, 2010) and aligns with suggestions that tweeting reflects dissemination rather than impact (Thelwall, Tsou, Weingart, Holmberg, & Haustein, 2013). The news/blog and Wikipedia results and to some extent the varying Twitter/GooglePlus/Facebook roles support previous claims that altmetrics might partly reflect non-scholarly impacts. Nevertheless, the associations are quite weak and the isolation of Wikipedia weakens the case for it reflecting a coherent dimension of impact.

The regression results show, for the first time, that Altmetric.com scores during the publication year can predict future citation counts in almost all fields (29 out of 30) assuming system stability. The $R^2$ scores were comparable to the prior regression of a medical journal's citations based on tweets and time, explaining 27% of variance (Eysenbach, 2011), although after excluding Mendeley, the Tweets and other metrics could account for more than 9% of the variance in only four cases (Table 6). Thus, the Journal of Medical Internet Research seems to be an anomalous case. The predictive power of the Altmetric.com scores declines substantially if Mendeley reader counts are excluded (Table 6), giving this reduced set marginal value in most fields.

## 4.1  *Journal Impact Factors as an alternative future citation predictor*

A scientist wishing to estimate their future citation counts might use an impact factor calculation from WoS or Scopus instead of Altmetric.com scores. Impact factors are controversial because they substitute document-level properties with publishing journal properties (DORA, 2013) and may be dominated by individual highly cited articles (Seglen, 1997). This need not rule out their use unless journals are believed to publish articles of heterogeneous quality and citations are believed to be accurate measures of quality (Waltman & Traag, 2017). At the individual article level, citation counts tend to correlate positively with journal impact factors, although the relationship has weakened since the 1990s (Lozano, Larivière, & Gingras, 2012).

Prior research shows that predicting later citation counts from early citations (e.g., one year) is possible in most cases (years 1-2 correlating with years 3-10 in six fields: Adams, 2005; year 1 citations correlating positively with 31 years of citations in 6 broad fields: Wang, 2013) but not all (Mingers, 2008). The predictions are very weak in most cases (e.g., Spearman correlations of 0.17-0.30 indicating weak rank predictions in: Wang, 2013), but two years of citations can give moderately good rank order predictions in at least one case (economics and political science: Stern, 2014; see also: Bruns & Stern, 2016). Longer term citation prediction is likely to be less accurate due to differing rates of citation accumulation for different articles (e.g., van Raan, 2004). Combining journal impact factor information with article-level citations gives more accurate predictions, however (Levitt & Thelwall, 2011; Stegehuis, Litvak, & Waltman, 2015). Although not explicitly tested here, authors should also consider when in the year their article was published since publications early in the year have a substantial early citation advantage over articles published later in the same year.

The effectiveness of the impact factor approach was checked in the current paper by allocating each article the 2014 Scopus CiteScore of its publishing journal from (journalmetrics.scopus.com). CiteScore records the average number of citations per document published in a journal 2011-2013 using Scopus citation counts from 31 May 2015. This is like the WoS Journal Impact Factor (JIF) except that it uses a three-year citation window instead of two years and includes all document types in its denominator rather than journal articles alone. This was used instead of the WoS JIF because it is available for (almost) all journals in the Scopus data set used here. The 2014 CiteScore was used as the latest available at the time of the November 2015 Altmetric data. The median correlation between 2017 Scopus citations and 2014 scores for each field was 0.522, which is double the corresponding median correlation between Mendeley readers and Scopus citations (0.247 in Table 3; all articles). The CiteScore correlation was higher than for Mendeley readers in 29 of the 30 individual fields.

Regression equations were constructed to predict 2017 Scopus citations from the November 2015 Altmetric.com scores and 2014 Scopus CiteScore data for articles with a positive Altmetric score and in a journal with a 2014 CiteScore. Across the 30 fields, the median $R^2$ for this new data set was 5% for Altmetric.com scores (excluding Mendeley) as independent variables, 21% when Mendeley was added as an extra independent variable, and 32% when both Mendeley and CiteScore were added (Table 7). The increase in $R^2$ when adding CiteScore for individual fields varies between 3% (Atmospheric Science) and 26% (Energy Engineering and Power Technology). Thus, adding CiteScore improves predictive power. In contrast, predicting with CiteScore alone gives a median $R^2$ of 20%, so Altmetric.com data improves predictions made on impact factor type data alone. Adding Altmetric.com scores to CiteScore improves predictive power ($R^2$) between 4% (Equine) and 18% (Aging).

In summary, Altmetric data can help authors to predict the future citation counts of their articles best if they consider the impact factor of the publishing journal first. Both sources are weaker on their own.

Table 7. R² values from fitting linear regressions. The Altmetric.com scores are from November 2015 and the future citation counts are from October 2017 for Scopus articles from 2015; CiteScore 2014 values are from May 2015. The regressions **exclude** articles with no positive Altmetric.com scores or without a CiteScore 2014 value.

| Field | Articles | Altmetric without Mendeley | Altmetric with Mendeley | Altmetric, Mendeley, CiteScore | CiteScore only |
|---|---|---|---|---|---|
| Agronomy & Crop Sci | 1251 | 9% | 23% | 36% | 26% |
| History | 1532 | 5% | 24% | 35% | 27% |
| Aging | 1557 | 7% | 25% | 30% | 12% |
| Accounting | 542 | 4% | 23% | 35% | 25% |
| Bioengineering | 2061 | 15% | 22% | 43% | 34% |
| Analytical Chemistry | 1736 | 5% | 15% | 28% | 19% |
| Artificial Intelligence | 922 | 2% | 20% | 29% | 18% |
| Info. Systems & Man | 362 | 7% | 26% | 39% | 24% |
| Atmospheric Science | 2640 | 7% | 17% | 20% | 9% |
| Econ & Econo | 1964 | 5% | 23% | 32% | 19% |
| Energy Eng & Power | 187 | 2% | 20% | 46% | 38% |
| Aerospace Eng. | 303 | 3% | 10% | 31% | 24% |
| Ecological Modeling | 717 | 4% | 18% | 33% | 20% |
| App Microb & Biotec | 3099 | 21% | 34% | 42% | 27% |
| Biomaterials | 2198 | 2% | 18% | 36% | 28% |
| Algebra & Numb Th | 121 | 7% | 16% | 19% | 7% |
| Anatomy | 1391 | 8% | 26% | 36% | 26% |
| Histology | 1649 | 2% | 19% | 25% | 14% |
| Rehabilitation | 2697 | 8% | 22% | 34% | 23% |
| Behavioral Neurosci | 3513 | 4% | 19% | 22% | 8% |
| Adv & Spec. Nursing | 1651 | 26% | 41% | 60% | 51% |
| Research & Theory | 144 | 5% | 18% | 23% | 7% |
| Drug Discovery | 1883 | 6% | 10% | 19% | 12% |
| Acoustics & Ultra | 371 | 5% | 16% | 23% | 10% |
| Applied Psych. | 3359 | 8% | 23% | 33% | 20% |
| Archeology | 806 | 8% | 21% | 34% | 26% |
| Urban Studies | 854 | 7% | 27% | 31% | 14% |
| Equine | 258 | 2% | 7% | 20% | 16% |
| Oral Surgery | 995 | 5% | 17% | 26% | 16% |
| Chiropractics | 152 | 17% | 27% | 33% | 17% |

## 5   Conclusions

The predictive power of Altmetric.com scores for articles during the year in which they are published is substantial enough to allow researchers and managers to use them to estimate future research impact, although they should be aware that they are only approximate indicators and that they should include Mendeley reader counts. If Mendeley counts are to be ignored for policy reasons then the predictions will be weak except in special cases (Advanced & Specialized Nursing; Applied Microbiology & Biotechnology). The predictive power is weaker than that of journal impact factors, however. Thus, journal impact factors (e.g., CiteScore) available when an article is published are better predictors of its longer-term citation counts than current article-level alternative indicators. The best strategy is to consider both, however.

When aggregated over sufficiently many articles, non-scholarly dimensions of impact may also be detected from Altmetric.com scores, including newsworthiness and social web interest. Despite this, there is not a strong underlying factor that could reasonably be described as societal benefit for all fields and so altmetrics should not be relied upon generically for evidence of non-scholarly impacts. Nevertheless, there is some suggestion that this might be possible in some fields (when the Mendeley/citations axis is not the main factor in Table 2). This would require field-specific investigations to ratify since the existence of a clear non-scholarly factor does not imply that the factor reflects societal impact because it could also reflect educational impact or even casual/recreational interest.

Although this article has introduced statistical tests and regression equations to assess the informational content of Altmetric.com scores, its primary goal is to assess whether it is reasonable for practicing scientists to use Altmetric.com scores as a rough early guide to the likely longer-term impact of their articles. In this context, it would be unreasonable for them to use regression to do this but reasonable to consider Altmetric.com scores intuitively, in conjunction with journal impact, to get an idea of which articles are more likely to attract longer term citations. Whilst, in theory, Altmetric.com could introduce a weighted formula for their Altmetric.com score to give it the greatest predictive power, their current strategy of reporting exact values and a simple sum seems better for its clarity, given the relatively low level of extra value given by the formula. Finally, the statistical analyses reported in this article should not obscure the fact that citation counts are very rough guides to the intrinsic quality or value of academic publications. Within a field (excluding the arts and humanities) and year, more cited articles tend to be better (Franceschet & Costantini, 2011; HEFCE, 2015) but highly cited articles can be poor (Ioannidis, 2005) and little cited articles can be useful (MacRoberts & MacRoberts, 2010); expert judgement is a better guide to quality.

# 6   Acknowledgement

Thank you to Altmetric.com for sharing the data used in this article.